\pdfoutput=1

\documentclass[twoside,11pt]{article}
\usepackage[accepted]{melba}

\usepackage[usenames,dvipsnames]{color}

\usepackage{amsmath,amsfonts}
\usepackage{booktabs}
\usepackage[table]{xcolor}
\usepackage{etoolbox}
\AtBeginEnvironment{tabular}{\tiny}
\usepackage{array}

\usepackage{multicol}
\usepackage{multirow}
\usepackage{longtable}
\usepackage{tabularray}

\bgroup\catcode`\#=12\egroup \newcommand\ExtraSep
{\dimexpr\cmidrulewidth+\aboverulesep+\belowrulesep\relax}

\melbaheading{2022:021}{https://www.melba-journal.org/papers/2022:021.html}{2022}{1-35}{04/2022}{09/2022}{Zeidan, Ramirez Gilliland, Patel, Ou, Flouri, Mufti, Maksym, Aughwane, Ourselin, David and Melbourne}{PIPPI 2021 special issue}{Esra Abaci-Turk, Jana Hutter, Roxane Licandro, Christopher Macgowan, Andrew Melbourne}

\ShortHeadings{Approach to automated diagnosis \& texture analysis of fetal organs in FGR}{Zeidan et al}
\firstpageno{1}

\title{PIPPI2021: An Approach to Automated Diagnosis and Texture Analysis of the Fetal Liver \& Placenta in Fetal Growth Restriction}

\author{\name Aya Mutaz Zeidan* \email aya.zeidan@kcl.ac.uk \\  
	\addr School of Biomedical Engineering \& Imaging Sciences, King's College London, UK
	\AND
	\name Paula Ramirez Gilliland* \email paula.ramirez\_gilliland@kcl.ac.uk \\
	\addr School of Biomedical Engineering \& Imaging Sciences, King's College London, UK
	\AND
	\name Ashay Patel \email ashay.patel@kcl.ac.uk\\
	\addr School of Biomedical Engineering \& Imaging Sciences, King's College London, UK
	\AND
	\name Zhanchong Ou \email zhanchong.ou@kcl.ac.uk\\
	\addr School of Biomedical Engineering \& Imaging Sciences, King's College London, UK
	\AND
	\name Dimitra Flouri \email dimitra.flouri@kcl.ac.uk\\
	\addr School of Biomedical Engineering \& Imaging Sciences, King's College London, UK
	\AND
	\name Nada Mufti \email nada.mufti@kcl.ac.uk\\
	\addr Elizabeth Garrett Anderson Institute of Women's Health, University College London, UK\\
	School of Biomedical Engineering \& Imaging Sciences, King's College London, UK
	\AND
	\name Kasia Maksym \email k.maksym@ucl.ac.uk \\
	\addr Elizabeth Garrett Anderson Institute of Women's Health, University College London, UK
	\AND
	\name Rosalind Aughwane \email r.aughwane@ucl.ac.uk \\
	\addr Elizabeth Garrett Anderson Institute of Women's Health, University College London, UK
	\AND
	\name Sebastien Ourselin \email sebastien.ourselin@kcl.ac.uk \\
	\addr School of Biomedical Engineering \& Imaging Sciences, King's College London, UK
	\AND
	\name Anna David \email a.david@ucl.ac.uk \\
	\addr Elizabeth Garrett Anderson Institute of Women's Health, University College London, UK
	\AND
	\name Andrew Melbourne \email andrew.melbourne@kcl.ac.uk \\
	\addr School of Biomedical Engineering \& Imaging Sciences, King's College London, UK\\
	Department of Medical Physics and Biomedical Engineering, University College London, UK
	\AND
	\addr *The first two authors contributed equally.
}

\begin{document}

\maketitle

\begin{abstract}
Fetal growth restriction (FGR) is a prevalent pregnancy condition characterised by failure of the fetus to reach its genetically predetermined growth potential. The multiple aetiologies, coupled with the risk of fetal complications - encompassing neurodevelopmental delay, neonatal morbidity, and stillbirth - motivate the need to improve holistic assessment of the FGR fetus using MRI. We hypothesised that the fetal liver and placenta would provide insights into FGR biomarkers, unattainable through conventional methods. Therefore, we explore the application of model fitting techniques, linear regression machine learning models, deep learning regression, and Haralick textured features from multi-contrast MRI for multi-fetal organ analysis of FGR. We employed T2 relaxometry and diffusion-weighted MRI datasets (using a combined T2-diffusion scan) for 12 normally grown and 12 FGR gestational age (GA) matched pregnancies (Estimated Fetal Weight below 3rd centile, Median 28$^{+4}$wks$\pm$3$^{+3}$wks). We applied the Intravoxel Incoherent Motion Model, which describes circulatory properties of the fetal organs, and analysed the resulting features distinguishing both cohorts. We additionally used novel multi-compartment models for MRI fetal analysis, which exhibit potential to provide a multi-organ FGR assessment, overcoming the limitations of empirical indicators - such as abnormal artery Doppler findings - to evaluate placental dysfunction. The placenta and fetal liver presented key differentiators between FGR and normal controls, with significant differences in features related to decreased perfusion, abnormal fetal blood motion and reduced fetal blood oxygenation. This may be associated with the preferential shunting of the fetal blood towards the fetal brain, affecting supply to the liver. These features were further explored to determine their role in assessing FGR severity, by employing simple machine learning models to predict FGR diagnosis (100\% accuracy in test data, n=5), GA at delivery, time from MRI scan to delivery, and baby weight. Moreover, we explored the use of deep learning to regress the latter three variables, training a convolutional neural network with our liver and placenta voxel-level parameter maps, obtained from our multi-compartment model fitting. 
Image texture analysis of the fetal organs demonstrated prominent textural variations in the placental perfusion fractions maps between the groups (p$<$0.0009), and spatial differences in the incoherent fetal capillary blood motion in the liver (p$<$0.009).
This research serves as a proof-of-concept, investigating the effect of FGR on fetal organs, measuring differences in perfusion and oxygenation within the placenta and fetal liver, and their prognostic importance in automated diagnosis using simple machine learning models.

\end{abstract}

\begin{keywords}
  Fetal Growth Restriction, Logistic Regression, Convolutional Neural Network, Texture Analysis
\end{keywords}

\section{Introduction}
The term Fetal Growth Restriction (FGR) is used to describe a fetus that has not reached their genetic growth potential, due to placental insufficiency causing inadequate supply of oxygen and nutrients (\cite{lyall2013spiral}). FGR is a clinical diagnosis, defined by the Delphi consensus standardised definitions (\cite{Gordijn2016}), and is divided into two different phenotypes, with onset either early (less than 32 weeks gestational age (GA)) or late in gestation. It is associated with high rates of stillbirth (\cite{Gardosi2013}), and neonatal morbidity including increased rates of cerebral palsy, bronchopulmonary dysplasia, and cardiovascular disease long term (\cite{Colella2018}). There is currently no treatment for FGR, therefore clinicians must weigh the risks of prematurity against the risk of hypoxia and death in utero to determine the optimal delivery time. There are limited clinical tools to do this, so at present, clinicians follow national guidelines to make this decision (\cite{no2002investigation}).

Considering the complicated nature of treatment and management, understanding the role and development of each organ during FGR is key for effective diagnosis and patient-specific severity assessment of the condition. Studies up to date only include quantitative analysis of a single fetal organ, most commonly the placenta, fetal brain, and fetal liver (\cite{salavati2019possible,malhotra2017detection,miller2016consequences,chang2006predicting,ebbing2009redistribution}). Our research overcomes these limitations by incorporating a multi-organ analysis for FGR assessment from MRI scans.

MRI is increasingly used to image the placental circulation. The Diffusion-rElaxation Combined Imaging for Detailed Evaluation (DECIDE) multi-compartment model separates fetal and maternal flow characteristics of the placenta allowing measurement of the relative proportions of vascular spaces (\cite{melbourne2019separating,couper2020}). When applied in early-onset FGR, it identified reduced feto-placental blood oxygen saturation, where the degree of abnormality correlated with disease severity defined by ultrasound fetal and maternal arterial Doppler findings (\cite{Aughwane2020}). 

The motivation for this research was to compare MR derived parameters relating to perfusion and oxygenation within the placenta and three fetal organs (the brain, liver and lungs) between normally grown pregnancies and those complicated by early-onset FGR, through multi-compartment models and texture analysis.  This research serves as a preliminary investigation into statistical methods leveraging multi-contrast MRI techniques to identify FGR predictors and thereby predict FGR, its severity, and resulting clinical complications. We propose a set of standardised imaging tools, important features, and initial statistical approaches for use in larger studies. Distinguishing features were then used to predict FGR diagnosis and GA at delivery via simple machine learning models.

\section{Related Works}

\subsection{Single- \& Multi-Compartment Models}
Blood oxygenation level-dependent (BOLD) contrast is a \(T^{*}_{2}\)-weighted sequence. It is affected by variations in concentration of vascular oxygentation in the blood volume and magentic field inhomogeneities. Quantifying $T_2^*$ enables the determination of oxygen saturation by leveraging the relationationship between $T_2^*$ and deoxyhemoglobin (\cite{sinding2016placental, sinding2017prediction}). In FGR pregnancies, the placenta is hypoxic, displaying a reduced $T_2^*$ value which can be used as an FGR biomarker (\cite{ robinson1998magnetic, jiang2013blood}). Despite the potential of BOLD-MRI in measuring oxygen saturation, its use has not yet been validated in diagnosis of FGR and interpretation of the placental BOLD signal is complicated by several factors that influence changes in the signal (\cite{sinding2018placental, uugurbil2000magnetic,chalouhi2014bold,sorensen2015placental,turk2020placental}). Considering this, and due to the requirements of a gradient echo acquisition,  $T_2^*$ relaxometry is not quantified in the current research. 

Instead, T$_2$ relaxometry provides structural, functional, and morphological tissue information as T$_2$ transverse relaxation times depend on several factors encompassing water binding, macromolecular concentration, and most importantly, blood oxygenation levels (\cite{derwig2013association,saini2020normal}). Previous literature has shown that the placental T$_2$ times in SGA or FGR pregnancies are reduced with respect to normal pregnancies (\cite{derwig2013association}). T$_2$ relaxation times have been used to assess placental function in various applications (\cite{melbourne2019separating,melbourne2016placental,jacquier2021multi,stout2021quantitative}).
\par

This study extended on previous placental research by producing T$_2$ maps for different fetal organs. It was hypothesised that because of the brain-sparing effect, certain organs would have lower oxygen levels in FGR compared to healthy pregnancies, and thus reduced T$_2$ measurements would be extracted from blood flowing through non-prioritised organs in FGR pregnancies. Portnoy \textit{et al.} demonstrated the precise relationship between blood T$_2$ relaxation times and oxygen saturation (\cite{portnoy2017relaxation}) by making use of the Luz-Meiboom model, given by Equation \ref{hematocrit1}, presenting the exponential relationship,
\begin{equation}
\label{hematocrit1}
    R_2 = Hct\, R_{2,ery} + (1-Hct)\, R_{2,plas} + R_{2,ex} , 
\end{equation}
where $R_2 = \frac{1}{T_2}$, $R_{2,ery}$ is the erythrocyte (red blood cell) relaxation rate that depends on oxygen saturation, $Hct$ is the hematocrit (proportion of red blood cells in blood), and $R_{2,plas}$ is the plasma relaxation rate. \par

Diffusion-weighted (DW) MRI is a valuable method for investigating the fetal brain-sparing effect; providing measures of brain maturation and detection of brain lesions (\cite{arthurs2017diffusion}). This is attained by measuring water diffusion, which yields corresponding apparent diffusion coefficient (ADC) values. Arthurs \textit{et al.} established differences between healthy and severe FGR fetuses, frequently leading to the clinical decision of early delivery induction in the latter group (\cite{arthurs2017diffusion}). The time between the MRI examination and delivery was, on average, 7.69 weeks earlier for the FGR group compared to the healthy, thus highlighting the potential of DW-MRI for accurate diagnosis of growth restricted cases - allowing for appropriate management plans to be put in place.

Dynamic contrast-enhanced (DCE) MRI can spatially and quantitatively characterise maternal perfusion of placental insufficiency and tissue vasculature (\cite{ingram201853, schrauben2019fetal, frias2015using}). It describes the delivery of contrast agent to the maternal side and its transfer into the fetal blood pool in order to distinguish between individual vascular units of the placenta. DCE-MRI is the current gold standard for quantitative descriptions of vascular function (\cite{frias2015using, schabel2016functional}). Nonetheless, this technique has significant drawbacks as it requires an exogenous contrast. The clearance of contrast from the feto-placental system still requires further research. To that end, an imaging technique which does not include any safety concerns for the mother and fetus is more pertinent. 

Multi-compartment models refer to advanced mathematical models that separate the signal contributions from different tissue types (\cite{Aughwane2020}). Diffusion-relaxation models are growing in popularity and have found multiple applications such as in neuroimaging (\cite{kim2017diffusion}), and more recently in placental imaging, encompassing the assessment of placental function in FGR (\cite{melbourne2019separating,melbourne2016placental, hutter2019multi, jacquier2021multi}). A thorough overview of these techniques is provided in (\cite{slator2021combined}).

The DECIDE model identifies and separates the $T_2$ values corresponding to the fetal and maternal blood, enabling the quantification of fetal blood oxygen saturation. The precise mechanisms and assumptions describing the DECIDE and the Extended Intravoxel Incoherent Motion Model (IVIM) models are discussed in Section \ref{sec:model_fitting_methods}.

\subsection{Diagnosis Predictions using Machine Learning} 

Supervised Machine Learning (ML) refers to the employment of a predictive model with an assumed relationship between the input (features) and output (labels) variables. Its prominence in medical imaging has been significantly established in recent years, particularly in computer-aided diagnosis (\cite{erickson2017machine, giger2018machine}), due to the rise of `Big Data' and available computer power. Its contribution to ``intelligent imaging" is by virtue of its potential to advance and enhance detection and diagnosis of complex disorders, risk assessment, and therapy response (\cite{schoepf2007pulmonary, dundar2008multiple, summers2010improving, mitchell2008predicting}). The advantages of ML stem from its ability to draw connections and identify patterns between variables, surpassing human perception. However, its attribute as a `black-box function' makes it difficult to interpret the results from ML models and determine how features are used to arrive at predictions, thus ensuing in a lack of clinician trustworthiness in the models. Nonetheless, ML can be leveraged to assimilate information from datasets where the relationship between the input and output variables are unknown and to select the best features for a certain prediction. It can be used for \textit{decision support} by aiding clinicians in interpreting medical imaging findings rather than relying entirely on the model predictions alone.

ML enables the consolidation and unravelling of complex biomedical and healthcare data that overcomes the limitations of traditional statistical methods. The aim of these algorithms is to provide solutions to clinical problems by learning statistical associations of the features extracted from the images (\cite{shen2017deep}).

Current screening and diagnostic tools for FGR remain suboptimal (\cite{audette2018screening}). Delivery of improved clinical outcomes requires greater understanding of the multifactorial pathogenesis in early-onset FGR and distinguishing features or biomarkers of the condition (\cite{audette2018screening}). Analysis of a combination of multiple FGR indicators (\cite{gordijn2018building,gordijn2016consensus, beune2018consensus}), can be achieved through use of ML methods. Supervised ML models are increasingly being employed for early prediction and diagnosis of pregnancy conditions, including intrauterine growth restriction, pre-eclampsia, risk of stillbirth, preterm pregnancy, and gestational diabetes (\cite{crockart2021classification,burgos2020evaluation, caly2021machine, khatibi2021proposing, maric2020early, ye2020comparison, koivu2020predicting}).

Recent work conducted by (\cite{arabi2021prediction}) compared the performance of logistic regression and artificial neural networks in predicting overall and spontaneous preterm birth, on a dataset of 112,963 nulliparous women (singleton gestation) who delivered between 20-42 weeks gestation. The predictors included socio-demographic variables correlated with the risk of preterm birth, such as maternal age, income, education, race, folic acid use, etc. The prediction accuracy of both models in the first trimester was ambiguous. But by incorporating complications during pregnancy as additional predictors, the authors established a 20\% increase in the area under the curve (AUC) from the receiver operating characteristic curve (ROC) for artificial neural networks in the validation sample compared to the logistic regressor during the second trimester (80\% vs. 60\%). The prediction performance of this work cannot be directly compared to our study, given the substantial difference in sample size (being several orders of magnitude smaller), which greatly influences the statistical power of the study. Therefore, our study should be viewed only as a preliminary study, as gaining concrete and detailed information regarding model performance and generalisability, and the features driving each ML model cannot be easily extracted as in (\cite{arabi2021prediction}), where the statistical methods were applied to a much larger dataset.

Research into the prediction of stillbirth by (\cite{yerlikaya2016prediction}) employed a multivariate logistic regression analysis to deduce the contributions of varying maternal characteristics and medical history in stillbirth prediction. Correspondingly, Trudell \textit{et al.} generated models for the prediction of stilbirth using backward stepwise logistic regression (\cite{trudell2017stillbirth}). Both groups leveraged highly similar maternal demographics and medical history and concluded similar prediction performances ranging between 64\% to 67\% AUC.

Despite the comparable performance of conventional logistic regression and ML methods for diagnosis predictions in previous literature (\cite{yerlikaya2016prediction,trudell2017stillbirth,koivu2020predicting,ye2020comparison}), the former assumes linearity and independence between the features. As such, we extended our previous methods (\cite{zeidan2021texture}) which implemented logistic regression to diagnose FGR and assess its severity, to a convolutional neural network (CNN). Deep learning algorithms draw on higher-level features extracted from the lower-level features of input data (\cite{bengio2012deep}). In particular, the benefits can be observed in supervised learning due to the scalability of deep neural networks and feature learning abilities. The use of a CNN allows us to explore both spatial and intensity relationships at a voxel-wise level for each of our parameter maps - information which is otherwise excluded when employing simple logistic regression models over averaged maps. Thus, we aim to maximise feature extraction from our parameter maps via a CNN, where feature representation is more accurate to the underlying maps for each organ, as no high-level averaging takes place. Nonetheless, it is important to acknowledge that a considerable amount of data is crucial to obtaining robust ML models.

\section{Methods}
\begin{figure}
    \includegraphics[width=\textwidth]{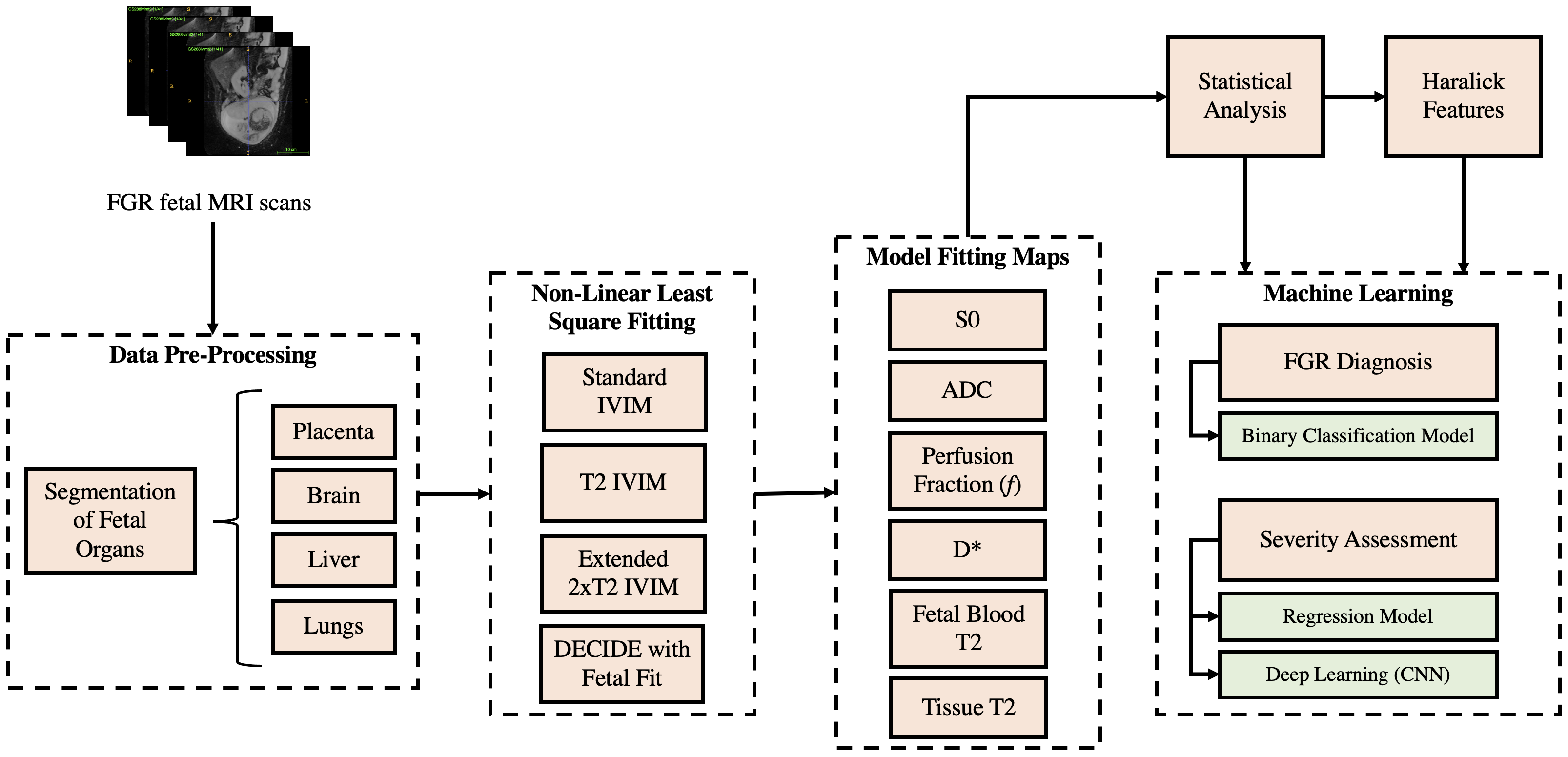}
    \caption{Overview of workflow. }
    \label{fig:outline}
\end{figure}

Model fitting techniques, described in Section \ref{sec:model_fitting_methods}, were applied to the segmented organs of interest, to yield quantitative parameters describing various signals. These parameters were then employed to perform texture analysis from multi-contrast MRI modelling, as described in Section \ref{sec:texture_analysis}. Results from the model fitting were used as inputs to the classifier and regressor in Sections \ref{sec:classification_methods} to \ref{sec:DL_methods} to predict a diagnosis of FGR and the GA at delivery. An overview of this pipeline is depicted in Figure \ref{fig:outline}.

\subsection{Data} \label{Data}

Patient MRI scans of voxel resolution 1.9x1.9x6$mm$ were acquired using the acquisition parameters from (\cite{melbourne2019separating}), where b-values and echo times are varied in pairs (enabling both T\(_{2}\) relaxometry and DW-MRI fitting), using a 1.5 T Siemens Avanto and performed under free-breathing. The dataset consisted of 12 early-onset FGR (\cite{Gordijn2016}) ranging between [24$^{+2}$, 33$^{+6}$] gestation weeks$^{+days}$, and 12 control pregnancies with MR data ranged between [25$^{+1}$, 34$^{+0}$] GA interval, (Median 28$^{+4}$wks$\pm$3$^{+3}$wks) respectively. Specific details on subject inclusion criteria are available in (\cite{Aughwane2020}). The study was approved by the UK National Research Ethics Service and all participants gave written informed consent (REC reference 15/LO/1488).

There are biological mechanisms that may cause differences in the distribution of blood perfusion throughout the fetus in FGR. To investigate this, manual segmentation of the placenta, liver, lungs and brain was accomplished using the open-source ITK-SNAP application (image segmentation). The resultant 3D mask files were used within the \textit{NiftyFit} package (\cite{melbourne2016niftyfit}) for multi-parametric model-fitting (\cite{melbourne2019separating}), and to perform texture analysis.  

\subsection{Model Fitting} 
\label{sec:model_fitting_methods}

Model fitting techniques were applied to each organ segmentation over the averaged region of interest (ROI) signal and on a voxelwise scale, yielding quantitative metrics for both approaches. Non-linear least squares were used to perform the fitting, with voxelwise fitting being initialised with the ROI parameter estimates - enhancing signal-to-noise ratio (SNR) by reducing the changes of fitting to local minima. A range of models were explored, including simple T\(_{2}\) and ADC estimation, as well as more complex models based on IVIM (\cite{le1986mr}) and DECIDE (\cite{melbourne2019separating}). Investigated in this research were parameters linked to diffusion, but they do not represent diffusion directly. The simplest T\(_{2}\) model fitting describes the MRI signal as  
\begin{equation}
\label{eq:t2_model}
    S = S_0 e^{-TE/T_2} , 
\end{equation}
where $TE$ are the echo times, $S$ is the measured signal, and $S_0$ the baseline signal. Regarding simple ADC fitting, this is accomplished using 
\begin{equation}
\label{eq:adc_model}
    S = S_0 e^{-b ADC} , 
\end{equation}
where b are the b-values. Thus, the acquired data requires varying $TE$ and b-values to allow for dual ADC and T2 model fitting. 

The IVIM model describes perfusion as a pseudodiffusion process (represented by a pseudodiffusion coefficient, D$^*$), by characterising the collective motion of blood water molecules within the vessel network as a random walk. The IVIM model also incorporates “true” diffusion of water molecules (ADC), modelling the signal as 
\begin{equation}\label{eq:DW_signal}
    S = S_0 [f e^{-b \text{D}^*} + (1-f) e^{-b \text{ADC}}] ,
\end{equation}
where $f$ is the perfusion fraction (volume occupied by incoherently flowing blood in a given voxel) and $b$ is the b-value (\cite{le2019can}). We refer to this model as \textbf{Standard IVIM} (Eq. \ref{eq:DW_signal}). This can be extended to incorporate T\(_{2}\) relaxometry as 
\begin{equation}
\label{eq:T2_ivim}
    S = S_0 e^{-t/\text{T}_2}[fe^{-b\text{D}^*}+(1-f)e^{-b \text{ADC}}] .
\end{equation}

We refer to this model (Eq. \ref{eq:T2_ivim}) as \textbf{T2 IVIM}. However, this model presents inherent limitations, as it assumes both vascular and tissue compartments (parametrised by pseudo-diffusion and true diffusion coefficients) have the same T\(_{2}\) value, leading to an overestimation of the pseudo-diffusion volume fraction \(f\) with increasing echo time ($t$) (\cite{jerome2016extended}). Thus, the presented analysis incorporates more complex models, accounting for varying blood and tissue T\(_{2}\) values:
\begin{equation}
\label{eq:extended_ivim}
    S(\mathbf{b,t}) = S_0[f e^{-b\text{D}^*} e^{-t/\text{T}_{2p}} + (1-f)e^{-b \text{ADC}}e^{-t/\text{T}_{2t}}] ,
\end{equation}
with \(f\) being the perfusion fraction, T\(_{2p}\) and T\(_{2t}\) being the transverse relaxation time for the pseudo-diffusion compartment (blood) and true diffusion compartment (tissue), respectively (\cite{jerome2016extended}). We refer to this model as \textbf{Extended \(2\times\)T2 IVIM} (Eq. \ref{eq:extended_ivim}). 

The \textbf{DECIDE} model (\cite{melbourne2019separating}) was also applied specifically to the placenta, which assumes three compartments with distinct diffusivity and relaxivity: fetal capillaries, trophoblast space and maternal blood pool. 
This model, given by Equation \ref{eq:DECIDE_model}, enables computation of novel placental biomarkers including maternal fetal blood volume ratio and fetal blood saturation.
\begin{equation}
\label{eq:DECIDE_model}
    S(\mathbf{b,t})\ =\ S_0\ [fe^{-b\text{D}^*-t (\text{1/T}_{2}^{fb})}+\left(1-f\right)e^{-b \text{ADC}\ \ }(\nu e^{-t (\text{1/T}_2^{mb})}+\left(1-\nu\right)e^{-t(\text{1/T}_2^{ts})})] .
\end{equation}

Here, T$_2^{fb}$, T$_2^{mb}$ and T$_2^{ts}$ represent the transverse relaxation times for fetal blood, maternal blood and trophoblast space, respectively; and $\nu$ is the maternal blood volume fraction. R$_2^{mb}$ and R$_2^{ts}$ are fixed known, $(240ms)^{-1}$ and $(46ms)^{-1}$ respectively at 1.5T), taken from (\cite{melbourne2019separating}).
\par

\subsection{Texture Analysis}
\label{sec:texture_analysis}

The aim of texture analysis was to examine the spatial arrangement of intensities in the segmented organs using in-house software developed in MATLAB (The MathWorks Inc., Natick, MA). To perform the texture analysis, a grey level co-occurrence matrix (GLCM) was computed to provide insight into the spatial interaction of neighbouring pixels. Haralick features are statistical features extracted from the GLCM to describe the overall image texture using measures encompassing energy, entropy, correlation, contrast, variance, and homogeneity (\cite{haralick1973textural}):

\textbf{\textit{Energy:} This measure is extracted from the angular second moment, which calculates the grey level local uniformity, 
\begin{equation}\label{eq:energy}
Energy = \sqrt{\sum_{i} \sum_{j} p^{2}_{d}\ (i, j)}
\end{equation}
where \(i\) and \(j\) represent the image dimensions, and \(p_{d}(i,j)\) corresponds to an element of the normalised GLCM.}

\textbf{\textit{Entropy:} A statistical measure of randomness.
\begin{equation}\label{eq:entropy}
Entropy = - \sum_{i} \sum_{j} p_{d}\ (i, j)\ ln\ p_{d}\ (i, j)
\end{equation}}

\textbf{\textit{Correlation:} A measurement of the similarity between neighbouring pixels,
\begin{equation}\label{eq:correlation}
Correlation = \sum_{i} \sum_{j} p_{d}\ (i, j) \frac{(i\ -\ \mu_{x}) (j\ -\ \mu_{y})}{\sigma_{x} \sigma_{y}}
\end{equation}
where \(\mu_{x}\); \(\mu_{y}\) are the means and \(\sigma_{x}\); \(\sigma_{y}\) are the standard deviations.}

\textbf{\textit{Contrast:} The number of grey levels that exist in the scan.
\begin{equation}\label{eq:contrast}
Contrast = \sum_{i} \sum_{j} (i\ -\ j)^{2}\ p_{d}\ (i, j)
\end{equation}}

\textbf{\textit{Variance:} A measure of variability.
\begin{equation}\label{eq:variance}
Variance = \sum_{i} \sum_{j} (i\ -\ \mu)^{2}\ p_{d}\ (i, j)
\end{equation}}

\textbf{\textit{Homogeneity:} The number of changes of intensity that appear in a region of interest. 
\begin{equation}\label{eq:homoegeneity}
Homogeneity = \sum_{i} \sum_{j} \frac{1}{1\ +\ (i\ -\ j)^{2}}\ p_{d}\ (i, j)
\end{equation}}

We hypothesised that these six Haralick features could be used to discern between FGR and appropriately grown fetuses due to a lower SNR present in FGR fetuses as a result of lower \(T_{2}\) and decreased oxygen saturation (\cite{portnoy2017relaxation}). We expected that the lower signal intensities in FGR compared to controls would be especially evident in the placenta and fetal liver and correlate directly with placental insufficiency (\cite{aughwane2020placental,kessler2009fetal}). For instance, this would be reflected in the computed Haralick features by observing lower contrast values in FGR fetuses in comparison to the controls. Decreased contrast in the ROI would equate to an increase in homogeneity.

These features were computed for each subject on the most significant parameter maps for each organ (as determined by the t-tests described in Section \ref{statistical_methods} with a p-value cut-off of 0.05), as well as the b=0 volume with lowest echo time from the original IVIM T$_{2}$-weighted MRI scan; this yielded interpretable texture descriptors (\cite{haralick1973textural,bharati2004image}). The images were quantised into grey level bins of fixed equal width for between-subject texture feature value comparisons. Single-factor analysis of each feature was conducted between the FGR and control patients. Results from the texture analysis were then concatenated by considering the mean and max of each Haralick feature.

\subsection{Feature Statistical Significance}
\label{statistical_methods}

The model fitting maps provide voxelwise information for each of the parameters optimised for. We simplified this information by considering the mean, max, variance and mode of each of voxelwise map. This yielded reduced parameters to be used for subsequent classical ML-based assessments.  

We performed statistical analysis on these simplified model fitting parameters and on the Haralick features, in order to identify the most significant features in differentiating between the control and FGR cohorts. A Shapiro–Wilk test was used to confirm normality of the parameters obtained from the model fitting on a patient-by-patient basis to verify it was justifiable to run a t-test on them. In particular, the Shapiro-Wilk test was selected for its efficacy on small sample sizes. The test was run on the distribution of the model fitted parameters for each of the organ ROIs done over all of the samples split between the two cohorts.



T-tests were then carried out between the two cohorts for all the model fitted parameters, Haralick features, and organ ratio parameters. Results with p-value less than 0.05 indicated statistically significant differences between the control and FGR group means.

We used these significant parameters for training simple classical machine learning models on classification (control or FGR) and regression (GA at birth, time from scan until birth and baby weight), as detailed in the following sections. 

\subsection{FGR Biomarkers for Machine Learning Outcome Predictions}

Following these statistical tests, we aimed to explore the use of these significant features (p-value $<0.05$ in distinguishing between controls and FGR) as potential FGR biomarkers for severity assessments. For this, we conducted various ML training experiments, employing a binary classifier for diagnosis prediction (control or FGR), and simple regressors to predict GA at birth, time from scan until birth, and baby weight.

Our training experiments explored the most appropriate use of our data to achieve optimal results. For this, we trained each model first using exclusively model fitting data (mean, max, variance and mode of each of voxelwise map), followed by exclusive training using Haralick features, and finally combining both model fitting data and Haralick features. Only the features with a p-value$<0.05$ in differentiating between controls and FGR cohorts were employed.

\subsection{Binary Classification for FGR Diagnosis}
\label{sec:classification_methods}

We employed logistic regression for binary classification, using a stochastic average gradient (SGA) solver that supports the L1 regularisation to minimise the cross-entropy loss function.

This classifier models the conditional probability of an FGR or non-FGR (control) diagnosis, \(\textbf{Y}\), 
given input features, \(\textbf{X}\) (model fitting data and Haralick features), by applying a sigmoid function to the output of a decision function \(h(\textbf{x}) = \textbf{w}^T \textbf{X}\), which ensures an output between 0 and 1:
\begin{equation}
\label{eq:MLmodels}
    P(\textbf{Y} = 1|\textbf{X}) = \frac{1}{1 + e^{- \textbf{w}^{T}\textbf{X}}} , 
\end{equation}

where \(\textbf{X}\) is the input feature vector, and \(\textbf{w} \) is the learnt weight vector. These probability scores (i.e. the output for Eq. \ref{eq:MLmodels}) are mapped to discrete classes with a decision boundary of 0.5, that is, an output probability\(<0.5 \) indicates an FGR diagnosis, while an output probability\(\geq 0.5 \) specifies a non-FGR diagnosis.

The optimal regularisation parameters (found via a grid search) were an L1 ratio of 0, i.e. L2 regularisation for all classifiers; and a regularisation strength (\(C\)) of 0.001 for the classifier trained exclusively on model fitting features, as well as the joint model (Haralick and model fitting features), while the model trained only on Haralick features yielded a \(C=0.25\).

Based on RFECV, we used 44 out of 84 features for the classifier trained on model fitting data; 34 out of 53 features for the classifier trained on Haralick features; and 118 out of 137 for the classifier trained on both feature types.

\subsection{Linear Regression Model for Severity Assessment}
\label{sec:regression_methods}

We trained three multi-variate linear regressors to predict GA at delivery, time interval between scan and delivery, and baby weight, as these variables (\(\hat{\textbf{y}} \)) are potential indicators of FGR severity. Thus we fitted a linear equation  \(\hat{\textbf{y}} = \textbf{X} \textbf{w} \) to our feature matrix \(\textbf{X}\), minimising the sum of squared errors between predicted and expected target values (including L1 and L2 regularisation) in order to find the weights, \(\textbf{w} \).

Refer to Table \ref{tbl:regressor} in Section \ref{regression_results} for information regarding model tuned hyperparameters and number of selected features for each regressor. 

\subsubsection{Training Split and Feature Selection}

For our simple ML classifier and regressors, the data was split into 80\% for training (n=18) and 20\% for testing (n=5). The training set was used for hyperparameter tuning using 5-fold cross validation. The reduced sample size in our research was confronted by additionally employing our training set to obtain 5-fold cross validated evaluation metrics, as well as the final test set metrics.

Recursive feature elimination with 5-fold cross validation (RFECV) was implemented on the training set to determine the optimal number of features for each ML model.

\subsection{Deep learning for Severity Assessment}
\label{sec:DL_methods}

The regression methods described in Section \ref{sec:regression_methods}, use data which statistically shows differences between FGR and healthy (p-value\(<\)0.05), followed by RFECV feature selection, to further reduce the noise present. However, the features used (in Section \ref{sec:regression_methods}), particularly the model fitting features, drastically reduce the amount of parameter maps information: by taking single statistical values over whole voxelwise maps (i.e. mean, max, min, mode), important spatial relationships and detailed voxel-level information may be eliminated. The Haralick features do contain information regarding spatial arrangements and intensity relationships, which supports our previous method.

In an attempt to make use of this detailed information contained within each parameter map, we explored the potential of a Convolutional Neural Network (CNN) for severity assessment, aiming to predict the same regression variables as with our simple ML models (GA at birth, time interval from scan until delivery, and baby weight). 

\subsubsection{Data pre-processing}

We used the voxelwise parameter maps for the liver and placenta only, as we found these organs to consistently have the highest number of significant differences between controls and FGR. We concatenated the first layer, the $S_0$ signal, of the volumetric image (i.e. the b=0 volume with the lowest echo time from the raw acquisition) with fitted parameter maps from the Extended T2 IVIM model for the liver (perfusion fraction \(f\), \(D*\), T\(_{2p}\), T\(_{2t}\) and ADC); and fitted parameter maps from the DECIDE model for the placenta (\(f\),  \(D*\), \(\nu\), T\(_{2fb}\), T\(_{mb}\), ADC), yielding a total of twelve input channels. 

The maps we selected were only those pertaining to the models which give us most information, which are the most complex models. Higher complexity models are more prone to add noise to the fitted maps. Contrasting to this, our input to the simple linear regressors were highly processed and selected features: we first took various extremely simplifying metrics of our voxelwise map (e.g. taking the mean), followed by selecting only those that present a statistical significance between both cohorts, in addition to RFECV. This provides us with features which are highly representative of distinctions between controls and FGR groups.

The data was split into 80\% for training (N=18), and 20\% for testing (N=5). The intensity of all images were normalised by subtracting the mean and dividing by the standard deviation, followed by scaling between 0 and 1. We used Gaussian noise, intensity shifts, bias field, contrast adjustments, axis flips, and affine deformations for data augmentation.

\subsubsection{CNN implementation details}

A five layer residual neural network (ResNet) (\cite{he2016deep}) was employed for each of our regression predictions, with output channels = [64, 64, 128, 256, 512], with respective strides = [1, 1, 2, 2, 2], applying two 3D convolutions for each residual block (kernel size of 3). Instance normalisation (\cite{ulyanov2016instance}) was used after each convolution, followed by Parametric Rectified Linear Unit (PReLU) activation functions (\cite{he2015delving}).

Mean Squared Error (MSE) was leveraged as the loss function, with an AdamW optimiser. A weight decay of \(5 \times 10^{-4}\) was employed for all of our regression networks, with a learning rate (LR) = \(5 \times 10^{-4}\) for predicting baby weight; and LR = \(5 \times 10^{-5}\) for predicting GA at birth and time from MRI scan to delivery.

\section{Results}

\subsection{Model Fitting}

\begin{figure}[htb]
    \centering
    \includegraphics[width=\textwidth]{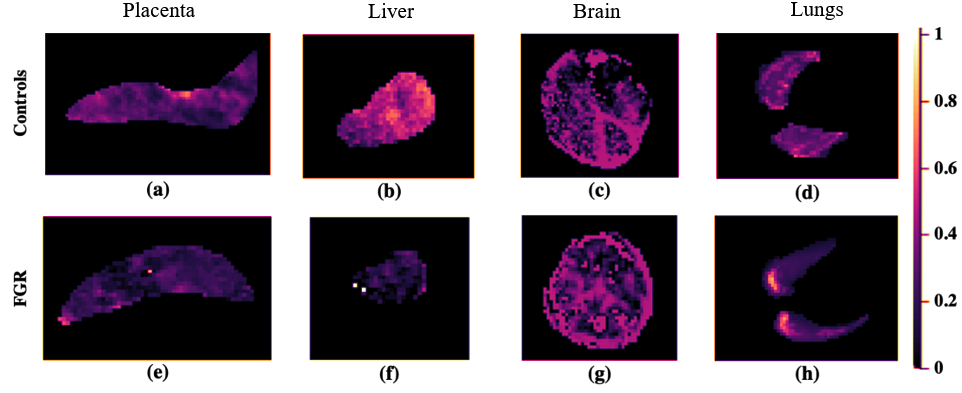}
    \caption{Perfusion fraction (dimensionless) layer in the model fitting maps each taken from a single slice in the MRI scan. These correspond to \textbf{((a),(e))} placenta, \textbf{((b),(f))} liver, \textbf{((c),(g))} brain and \textbf{((d),(h))} lungs. Top and bottom rows correspond to controls and FGR, respectively.}
    \label{fig:perfusion_maps_organs}
\end{figure}

Figure \ref{fig:perfusion_maps_organs} depicts examples of the parameter maps obtained from the model fitting techniques. The lower parameter map intensities in FGR compared to that in the controls is indicative of hypoperfusion and low oxygen saturation levels in these fetal organs. The T$_{2}$ maps display pronounced differences in the signal intensities of both cohorts.

\begin{figure}[htb]
    \centering
    \includegraphics[scale=0.22]{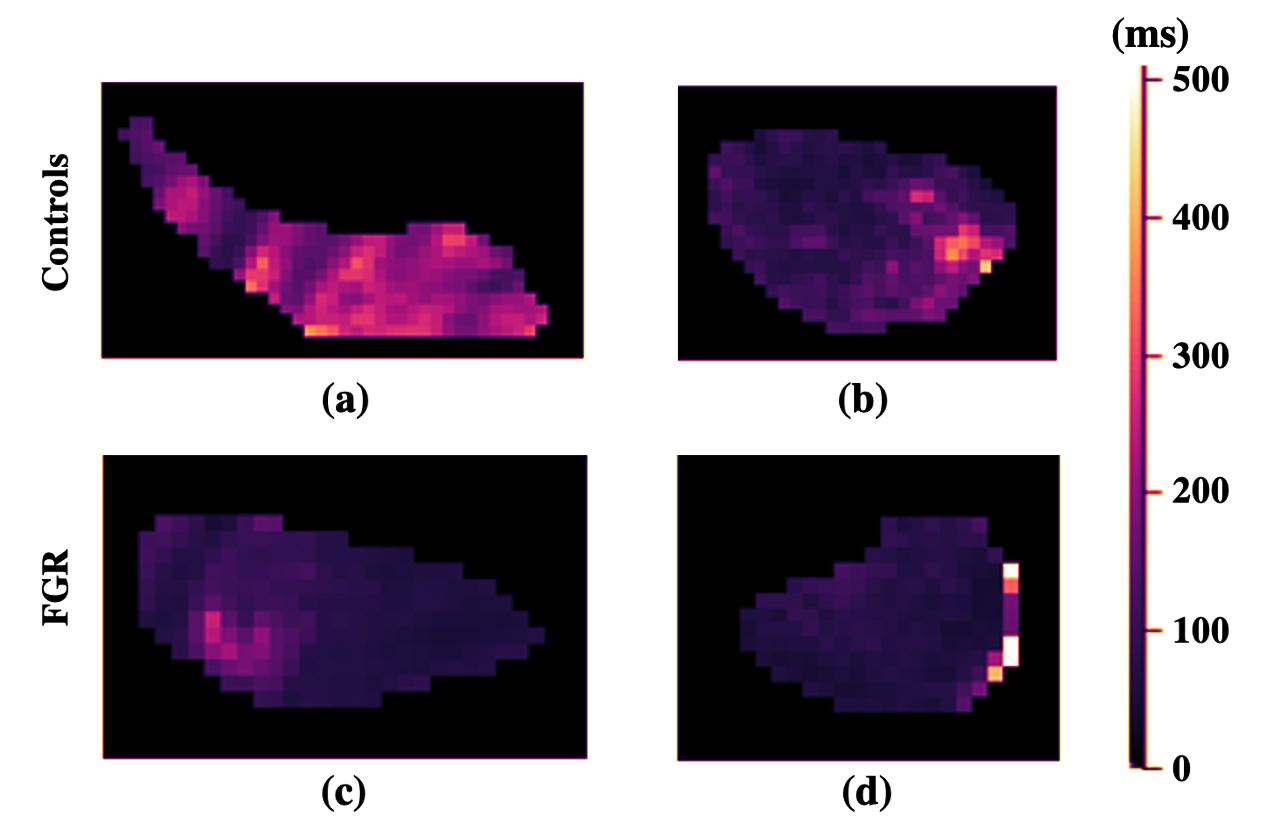}
    \caption{T$_{2}$ (units of \textit{ms}) maps for \textbf{((a), (c))} placenta, and \textbf{((b), (d))} liver from a single slice. Top and bottom rows correspond to controls and FGR, respectively. }
    \label{fig:t2_organ_maps}
\end{figure}

The most significant parameters in identifying differences between controls and FGR fetuses were the perfusion fraction, S$_0$, pseudo-diffusion coefficient (D$^*$), and T$_{2}$ as given in Table \ref{tbl:voxel_significance}. The placenta and liver were determined to be the most influential organs in diagnosing FGR.

The results for the parameter feature importances in Table \ref{tbl:voxel_significance}, specify that there were no significant differences detectable in the fetal brain and lungs between normal and FGR fetuses, especially compared to the placenta and liver, where differences were significant.



\begin{table}[!ht]
    \centering
    \begin{tabular}{|p{3.3cm}|p{1.3cm}|p{1cm}|p{2.5cm}|p{1cm}|p{1.5cm}|p{1.5cm}|}
    \hline
        \textbf{Model Fitting Technique} & \textbf{Parameter} & \textbf{Average Metric} & \textbf{Pairwise Group Comparison} & \textbf{Organ} & \textbf{T Statistic} & \textbf{P-Value} \\ \hline\hline
        Dependent IVIM & D* & Mean & Control vs FGR & Placenta & -4.597300242 & 0.00015589 \\ \hline
        Extended 2xT2 Dependent IVIM & D* & Mean & Control vs FGR & Placenta & -4.560436097 & 0.000170214 \\ \hline
        DECIDE Model (Voxelwise Measurements) & D* & Mean & Control vs FGR & Placenta & -4.205788361 & 0.00039723 \\ \hline
        Extended 2xT2 Dependent IVIM & Perfusion Fraction & Min & Control vs FGR & Placenta & 3.725183003 & 0.001250966 \\ \hline
        Extended 2xT2 Dependent IVIM & Perfusion Fraction & Mode & Control vs FGR & Placenta & 3.725183003 & 0.001250966 \\ \hline
        Standard IVIM & Perfusion Fraction & Median & Control vs FGR & Liver & 3.624757118 & 0.001587669 \\ \hline
        T2 Dependent IVIM & T2 & Min & Control vs FGR & Placenta & 3.463092031 & 0.002326109 \\ \hline
        Extended 2xT2 Dependent IVIM & Perfusion Fraction & Median & Control vs FGR & Placenta & 3.27041186 & 0.003653498 \\ \hline
        T2 Dependent IVIM & Perfusion Fraction & Min & Control vs FGR & Placenta & 3.249455242 & 0.003836258 \\ \hline
        T2 Dependent IVIM & Perfusion Fraction & Mode & Control vs FGR & Placenta & 3.249455242 & 0.003836258 \\ \hline
    \end{tabular}
    \caption{Hierarchy of parameter feature importances of the voxelwise fitted parameter map measurements. Refer to Appendix \ref{sec: voxelwise_appendix} for an extension of the table which includes the 50 most significant features.}\label{tbl:voxel_significance}
\end{table}

\subsection{Texture Analysis}

\begin{figure}[ht] 
\centering
\includegraphics[scale=0.45]{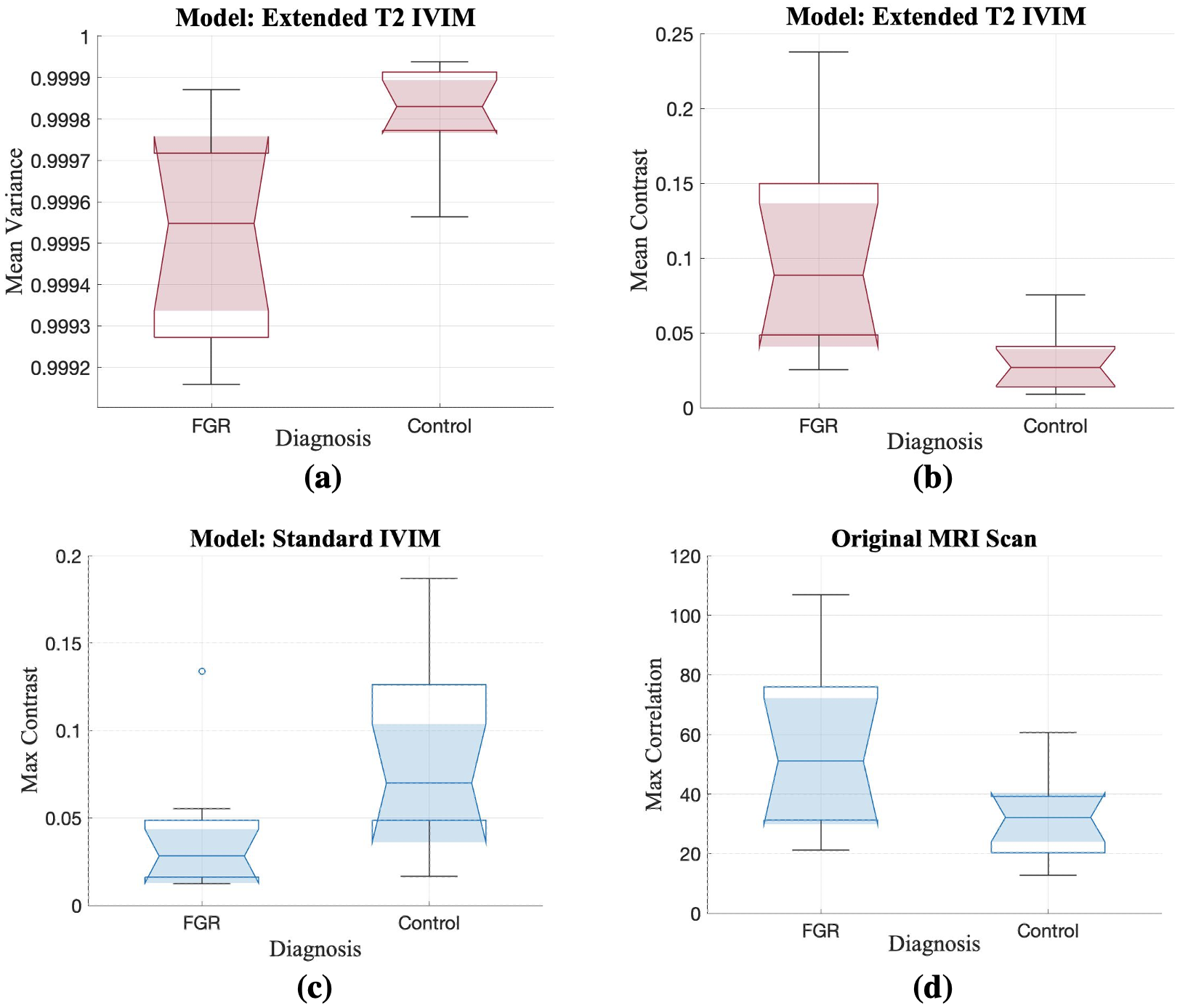}
\caption{\textbf{Comparison of most significant Haralick features.} Notched box plots of the most significant placental (pink) and liver (blue) Haralick features: {\textbf{(a, b)}} mean values of the variance and contrast of the perfusion fraction in the Extended T\(_{2}\) IVIM model, {\textbf{(c)}} max values of the contrast of the D* parameter in the Standard IVIM model, and \textbf{(d) max values of the correlation computed from the MRI scan (b=0 volume). (Refer to Appendix \ref{sec: haralick_appendix} for a breakdown of the 50 most significant Haralick features.) The notches in the box plots delineate the extent of significant difference in the medians of the investigated features by representing the confidence interval of the metric. }}
\label{fig:haralick_results}
\end{figure}

Evaluation of the resulting Haralick features corroborated the degree of effect on the placenta in FGR, particularly using the Extended T$_{2}$ IVIM map and its mean variance. The brain was the least significantly different organ in this analysis. Greater mean variance in the signal from the Extended T$_{2}$ IVIM model of the healthy cohort (refer to Figure \ref{fig:haralick_results}(a)), is indicative of increased heterogeneity in FGR placentas. The max correlation of the liver perfusion fraction in the controls in Figure \ref{fig:haralick_results}(d) reflects larger intensity differences compared to FGR. This is a significant feature to consider in the Standard IVIM model when studying the liver in FGR, especially given that the notches do not overlap between the cohorts.

\subsection{FGR Diagnosis via a Classification Model}
\label{classification_results}

\begin{table}[!ht]
    \centering
    \begin{tabular}{| p{2.5cm} | l | l l l | p{3cm} |}
    \hline
        \multicolumn{1}{|c|}{\multirow{2}{*}{\textbf{Training Dataset}}} & \multicolumn{1}{c|}{\textbf{Cross Validation (N = 18)}} & \multicolumn{3}{c|}{\textbf{Testing (N = 5)}} & \multicolumn{1}{c|}{\textbf{RFE}} \\\cline{2-6}\cline{2-6}
         ~ & \multicolumn{1}{c|}{\textbf{Accuracy}} & \multicolumn{1}{c}{\textbf{Accuracy}} & \multicolumn{1}{c}{\textbf{Sensitivity}} & \multicolumn{1}{c|}{\textbf{Specificity}} & \textbf{Top Five Features (Model)} \\ \hline
       \multirow{5}{*}{Model Fitting Features} & \multicolumn{1}{c|}{\multirow{5}{*}{95 ± 10\%}} & \multicolumn{1}{c}{\multirow{5}{*}{100\%}} & \multicolumn{1}{c}{\multirow{5}{*}{100\%}} & \multicolumn{1}{c|}{\multirow{5}{*}{100\%}} & Placenta mean D* (T2 IVIM) \\\cline{6-6}
        ~ & ~ & ~ & ~ & ~ & Placenta mean D* (Extended 2xT2 IVIM) \\\cline{6-6}
        ~ & ~ & ~ & ~ & ~ & Placenta mean D* (DECIDE) \\\cline{6-6}
        ~ & ~ & ~ & ~ & ~ & Liver/Lungs median perfusion fraction (Standard IVIM) \\\cline{6-6}
        ~ & ~ & ~ & ~ & ~ & Placenta/Lungs median perfusion fraction (Extended 2xT2 IVIM) \\\hline
        \multirow{5}{*}{Haralick Features} & \multicolumn{1}{c|}{\multirow{5}{*}{77 ± 12\%}} & \multicolumn{1}{c}{\multirow{5}{*}{80\%}} & \multicolumn{1}{c}{\multirow{5}{*}{67\%}} & \multicolumn{1}{c|}{\multirow{5}{*}{100\%}} & Placenta mean variance D* (Extended 2xT2 IVIM) \\\cline{6-6}
        ~ & ~ & ~ & ~ & ~ & Placenta max correlation D* (Extended 2xT2 IVIM) \\\cline{6-6}
        ~ & ~ & ~ & ~ & ~ & Placenta mean correlation D* (T2 IVIM) \\\cline{6-6}
        ~ & ~ & ~ & ~ & ~ & Liver max contrast D* (Standard IVIM) \\\cline{6-6}
        ~ & ~ & ~ & ~ & ~ & Liver mean contrast D* (Standard IVIM) \\\hline
        \multirow{5}{*}{Combined Features} & \multicolumn{1}{c|}{\multirow{5}{*}{88 ± 15\%}} & \multicolumn{1}{c}{\multirow{5}{*}{100\%}} & \multicolumn{1}{c}{\multirow{5}{*}{100\%}} & \multicolumn{1}{c|}{\multirow{5}{*}{100\%}} & Placenta mean D* (T2 IVIM) \\\cline{6-6}
        ~ & ~ & ~ & ~ & ~ & Placenta mean D* (Extended 2xT2 IVIM) \\\cline{6-6}
        ~ & ~ & ~ & ~ & ~ & Placenta mean D* (DECIDE) \\\cline{6-6}
        ~ & ~ & ~ & ~ & ~ & Liver/Lungs median perfusion fraction (Standard IVIM) \\\cline{6-6}
        ~ & ~ & ~ & ~ & ~ & Placenta/Lungs median perfusion fraction (Extended 2xT2 IVIM) \\ \hline
    \end{tabular}
    \caption{\textbf{Classification Results.} Evaluation metrics for cross-validation and testing for each classifier, alongside the top five features selected by Recursive Feature Elimination (RFE).}\label{tab:classification_results}
    \label{tbl:classifier}
\end{table}

Referring to the results presented in Table \ref{tbl:classifier}, the classifier performs best when trained exclusively on model fitting data, achieving a prediction accuracy of 100\% in testing, and thus a precision and recall score of 1.0.

This is further validated by the cross validated accuracy on training set, with a standard deviation of only 10\% across folds, hinting at optimal model generalisability.

\subsection{Classification Feature Importance}
\label{sec: classification_features}
Given our optimal test set classification results (see Table \ref{tbl:classifier}), we qualitatively assess the most important features driving each classifier model. These were obtained via Recursive Feature Elimination (RFE). We obtained the exact same top five features for both the logistic regressor trained exclusively on model fitting data, and the logistic regressor trained on both Haralick features and model fitting data.

\begin{figure}[h]
    \centering
    \includegraphics[width=\textwidth]{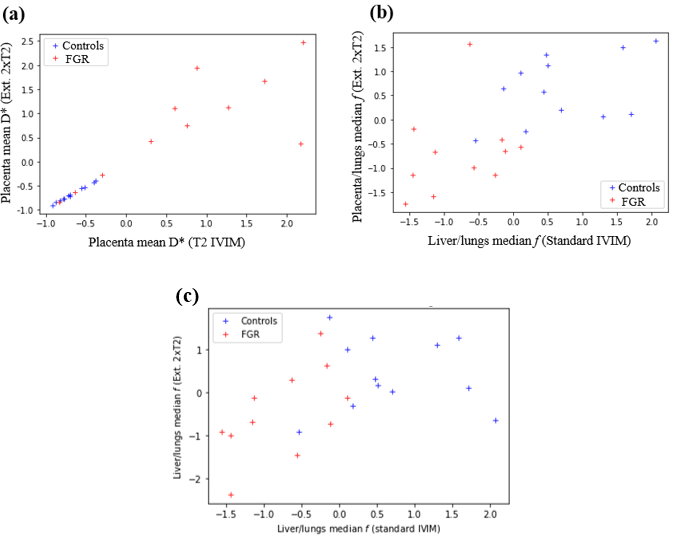}
    \caption{\textbf{Relationship between the features employed in the classification model.} (\textbf{a}): Placenta mean D* from the Extended 2\(\times\)T2 IVIM model against the T2 IVIM model. (\textbf{b}): Placenta/Lungs mean perfusion fraction from the Extended 2\(\times\)T2 IVIM model against Liver/Lungs median perfusion fraction from the Standard IVIM model. Note that values have been scaled. (\textbf{c}): Liver/Lungs median perfusion fraction from the Extended 2xT2 IVIM model against the Standard IVIM model.}
    \label{fig:classification_features}
\end{figure}

Figure \ref{fig:classification_features}\textbf{b} shows distinct differences between controls and FGR cohorts. Here, control subjects display a much higher placenta and liver perfusion relative to the lungs, compared to FGR subjects. This is an indicator that in FGR, both the placenta and the lungs are much less perfused than other vital organs such as the lungs. 

The differences between the performance of the model fitting techniques can be inferred from Figure \ref{fig:classification_features}\textbf{c}. It showcases the `liver/lungs median \textit{f}' plotting the Extended 2xT2 IVIM model against the Standard IVIM model. When projected onto each axis, the x-axis (i.e. the Standard IVIM model), permits a more accurate linear classification between the cohort compared to the Extended 2xT2 IVIM model. Seemingly, the Standard IVIM model produces a better fit to our data (for this particular parameter) than the more complex models with additional parameters, potentially due to the noise present.

\subsection{Severity Assessment via a Regression Model}
\label{regression_results}

\begin{table}[!ht]
    \centering
    \begin{tabular}{|p{1.6cm}|p{1.4cm}|p{2cm}|p{2cm}|p{2cm}|p{1.5cm}|p{1.5cm}|}
    \hline
        \multirow{3}{=}{\begin{tabular}{l} \textbf{Prediction}\end{tabular}} & \multirow{3}{=}{\begin{tabular}{l} \textbf{Training} \\ [\ExtraSep] \textbf{Dataset}\end{tabular}} & \multirow{3}{=}{\begin{tabular}{l} \textbf{Regularisation} \\ [\ExtraSep] \textbf{strength ($\alpha$)} \end{tabular}} & \multirow{3}{=}{\begin{tabular}{l}\textbf{Regularisation} \\ [\ExtraSep] \textbf{Ratio (L1/L2)}\end{tabular}} & \multirow{3}{=}{\begin{tabular}{l}\textbf{RFECV} \\ [\ExtraSep] \textbf{(Selected} \\[\ExtraSep] \textbf{Features/Total} \\ [\ExtraSep] \textbf{Features)}\end{tabular}} & \textbf{Cross Validation (N = 18)} & \textbf{Testing (N = 5)} \\ \cline{6-7}
        ~ & ~ & ~ & ~ & ~ & \textbf{RMSE ± STDEV} & \textbf{RMSE} \\
         ~ & ~ & ~ & ~ & ~ & ~ & ~ \\\hline
        \multirow{3}{*}{\shortstack{ GA at\\ [\ExtraSep]
        Delivery }} & Model Fitting Features & 33.93 & L2 only & 71/84 & \textbf{2.9 ± 2.36 weeks} & \textbf{2.1 weeks} \\ \cline{2-7}
        ~ & Haralick Features & 0.49 & L1 only & 5/53 & 4.48 ± 4.13 weeks & 3.06 weeks \\ \cline{2-7}
        ~ & Combined Features & 44.98 & L2 only & 119/137 & 3.0 ± 2.42 weeks & 3.1 weeks \\ \hline
        \multirow{4}{*}{\shortstack{
                   Time from\\  [\ExtraSep]
                   scan until\\  [\ExtraSep]
                   delivery}} & Model Fitting Features & 59.64 & L2 only & 84/84 & \textbf{3.21 ± 2.53 weeks} & 3.12 weeks \\ \cline{2-7}
        ~ & Haralick Features & 1.15 & L1 only & 5/53 & 4.95 ± 3.51 weeks & 4.82 weeks \\ \cline{2-7}
        ~ & Combined Features & 7.20x10$^{-3}$ & 0.31 & 133/137 & 3.5 ± 2.68 weeks & \textbf{3.09 weeks} \\ \hline
        \multirow{3}{*}{\shortstack{ Baby weight}} & Model Fitting Features & 2.32x10$^{-3}$ & 0.16 & 64/84 & \textbf{372.71 ± 334.42 g} & \textbf{991.36 g} \\ \cline{2-7}
        ~ & Haralick Features & 25.6 & 0.92 & 28/53 & 738.88 ± 600.58 g & 1591.72 g \\ \cline{2-7}
        ~ & Combined Features & 3.56 & L1 only & 5/137 & 668.64 ± 488.42 g & 1099.06 g \\ \hline
    \end{tabular}
    \caption{\textbf{Regression Results.} Tuned linear regressor model hyperparameters with corresponding evaluation metrics for cross-validation and testing (RMSE). Best results for each outcome are highlighted in bold.}
    \label{tbl:regressor}
\end{table}

Table \ref{tbl:regressor} includes our test set and cross validated regressor results. In accordance with our classifier results, the models with highest performance are those trained on model fitting features, excepting predictions for time from scan until delivery, where the combined model displays an insignificantly lower root mean square error (RMSE) on test set compared to the model trained exclusively on model fitting data. 

\begin{figure}[h]
    \centering
    \includegraphics[width=\textwidth]{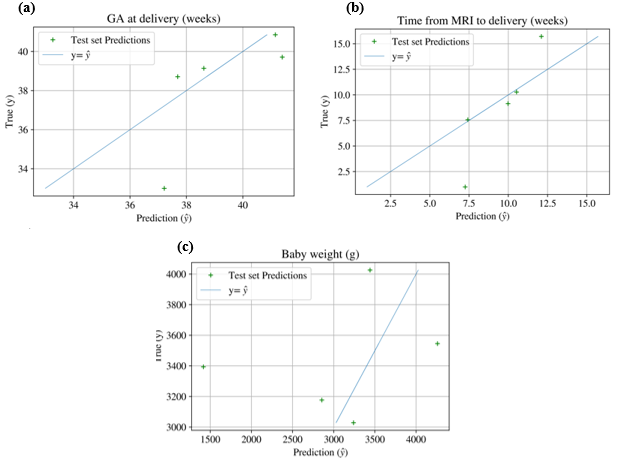}
    \caption{Regression test set results against true features for GA at delivery (\textbf{a}), time interval between MRI scan and delivery (\textbf{b}), and baby weight (\textbf{c}).}
    \label{fig:regression_true_vs_pred}
\end{figure}

Figure \ref{fig:regression_true_vs_pred} depicts test set regression predictions for our best performing regressors, against true labels. Qualitatively, the 
test set predictions that mostly resemble the true data points are from the time interval between MRI and delivery (Figure \ref{fig:regression_true_vs_pred}\textbf{b}), however this has two important outliers. It is complex to comment on the significance of this, given the extremely small test set. The plot depicting baby weight predictions (Figure \ref{fig:regression_true_vs_pred}\textbf{c}) visually appears as the worst fit, however the value range for this variable is much larger, which may partially explain this. Additionally, the most important outliers for baby weight are predictions which are lower than the actual baby weight, which is clinically significant: it is best to overestimate the severity than underestimate it.

\subsection{Deep Learning Regression}
\label{sec:deep_learning_results}

\begin{table}[!ht]
    \centering
    \begin{tabular}{|l|l|l|}
    \hline
        \multicolumn{3}{|c|}{\textbf{RMSE on Test Set (N = 5)}} \\ \hline
        \textbf{GA at delivery} & \textbf{Time from scan until delivery} & \textbf{Baby weight} \\ \hline
        5.33 weeks & 5.93 weeks & 1169.88 g \\ \hline
    \end{tabular}
    \caption{Results for ResNet regression predictions on test set.}\label{tbl:cnn_results}
\end{table}

The ResNet results for severity assessment are included in Table \ref{tbl:cnn_results}. Both GA at birth and time interval between scan and delivery resulted in a much higher RMSE than those obtained from our simple classical linear regressors (see Section \ref{regression_results}).

\section{Discussion}

In this study, we combined model fitting techniques, texture analysis from multi-contrast MRI modelling, and ML models, to facilitate multi-fetal organ analysis of FGR. This provided a more holistic approach to imaging this common pregnancy condition and presented an approach towards automated diagnosis and severity assessment. Differences between FGR and non-FGR fetuses were observed, particularly in the placenta and fetal liver, emphasising the significant effect of FGR on these organs.

Overall, the fitted model parameters reveal decreased \textit{f}, T$_{2}$, and D$^*$ in the liver and placenta in FGR fetuses compared to the controls. These findings are validated by those from (\cite{shi2019evaluation, siauve2019assessment, razek2019apparent, aughwane2021magnetic}). The hierarchy of feature importances in Table \ref{tbl:voxel_significance} suggests that the brain and lungs may benefit from alternative analysis, focusing on certain cortical regions for the brain, and incorporating alternative imaging modalities for the lungs, as model fitting MRI analysis may not be the most appropriate technique for this fluid-filled organ. These differences are indicative of a reduced oxygen saturation and perfusion within these organs, as well as abnormal capillary blood flow motion (\cite{Aughwane2020}). We did not observe significant differences in the properties of fetal brains and lungs between the FGR and control groups. 

The most influential Haralick features were extracted from the perfusion fraction measurements, particularly computed from the Extended T$_{2}$ IVIM and Standard IVIM models. Another important parameter determined by the Haralick features was T$_{2}$, attributed to its correlation with oxygen saturation (lower T$_{2}$ reflects a lower oxygen saturation (\cite{portnoy2017relaxation})). 

The placenta was established as the organ with most significant textural differences between the FGR and control groups. Variance, contrast, entropy and energy in placental perfusion fraction maps were the most significant textural differences between FGR and controls. This may be related to differences in the presence of maternal and fetal vascular malformation (\cite{Mifsud2014,Burton2009}). 

The second organ with greatest textural differences between both cohorts was the liver, particularly the D$^*$ maps (contrast, correlation, and energy), indicating spatial differences in the incoherent fetal capillary blood motion in this organ. This may indicate an abnormal blood motion in the liver compared to a healthy developing organ, affecting nutrient supply to this organ and may be related to the role of the ductus venosus in redistributing blood to the heart under the influence of increasing hypoxia (\cite{mifsud2014placental}). Energy was heavily influenced by the number of grey levels and was, therefore, a significant feature for the placenta, lungs and brain, due to the presence of similar intensity voxels within local regions. Correlation was affected by the noise present in the image, which explains the notable correlation differences found in the liver, being the organ with the lowest SNR. 
\par 

The feature importances determined by RFE for the classifier in Section \ref{sec: classification_features}, and the fact that they coincide with the top five features for the logistic regressor, indicate that these are very strong features in determining model predictions. The top five features from our best regressor models are those involving the liver \textit{f}, placenta \textit{f}, placenta D*, placenta tissue T2, and liver/lung D*. Thus, the top features we obtained here are very similar to those from our binary classifiers, which strengthens our argument that the liver and placenta may be less well perfused in FGR, with altered circulation patterns. We additionally found placental tissue T2 as a significant feature for severity predictions. Tissue T2 is related to tissue oxygenation. Therefore, the fact that this is one of our most important features for severity assessment may be linked to reduced placental oxygenation in more severe cases, affecting fetal growth.  

In particular, these features all involve either the placenta and the liver, which supports our prior t-tests and Haralick feature analysis. Two markedly informative features are the ratios of Placenta/Lungs and Liver/Lungs median perfusion fractions (\textit{f}). These features suggest that in control subjects, the relative perfusion of the liver and placenta compared to the lungs is much higher than in FGR cases, i.e. the liver and placenta are not deprived of nutrients, as may be the case in FGR.

The recurrence of the D* and \(f\) in the top features demonstrate these may be potential FGR biomarkers. Figure \ref{fig:classification_features} includes a visual depiction of the mean D*, as computed from two different models. The linear relationship between variables on the leftmost plot (\textbf{a}) is due to the axis representing the same variable, as computed from two different models, thus differences are due to different model assumptions and noise. This plot clearly show abnormal D* placenta values for FGR subjects, with these displaying a much larger spread compared to controls. This pseudo-diffusion coefficient (D*) describes macroscopic intra-capillary blood motion. Thus, these results are suggestive of abnormal placental circulatory patterns, which may be due to placental insufficiencies and dysfunctions in FGR. 

While blood in the intervillous space appears to undergo incoherent motion, the maternal blood fraction is not attributed to D* in addition to ADC in Equation \ref{eq:DECIDE_model}, as previously modelled by (\cite{melbourne2019separating}). Our working assumption within the modelling is that maternal blood arrives at high-flow, low-velocity, resulting in an overall lower D* value compared to that for fetal intra-capillary blood and moves slowly through the villous structure. It is probable that this assumption is less true close to the spiral artery inlets - but this remains to be fully validated.

Placental and liver perfusion fraction, D* and tissue T2 were amongst the most important features for our ML binary classifiers and linear regressors, as determined by RFE. This supports our choice of most important textural differences and aforementioned biological reasoning. The classifier achieved 100$\%$ accuracy on the test set, indicating that the model features are powerful indicators for FGR detection. But these results require prospective validation in a larger study population due to the small test group size (n=5) in this proof-of-concept study, which may have resulted in overfitting of the models to the features. Moreover, a larger dataset would permit the transition into more complex prediction models in future research. \par

The RMSE of 2.1 weeks and 3.09 weeks for our linear regressor predicting GA at delivery and time interval from scan until delivery, respectively, encode a large window in terms of fetal development. Recent research conducted by Yamauchi \textit{et al.} employed leave-one-out cross validation to predict GA in normal and complicated pregnancies from urinary metabolite information (\cite{yamauchi2021machine}). The authors achieved a Pearson correlation coefficient of 0.86 between the true and predicted GAs during normal pregnancy progression, and an RMSE of 26.7 gestation days (3.81 weeks). Thus, the performance of our regressor appears to be comparable with that from a model trained on 187 healthy pregnant women. \par

The results in Table \ref{tbl:regressor} indicate a lower RMSE from the combined model compare to the model trained exclusively on model fitting data. This signifies that our model fitting maps have a higher difference in intensity values, rather than textural or spatial relationships, between control and FGR cohorts, and for varying degrees of condition severity. From a mathematical point of view, considering that our data presents a range of 3864 g for baby weight, 15 weeks for GA at birth and 15.57 weeks for time interval between MRI scan and delivery; our RMSE on test set only represent 25\%, 19.85\% and 14\% out of our total dataset range for baby weight, scan to birth interval, and GA at birth, respectively.  

However, from a clinical perspective, offering a prediction with a RMSE of 2-3 weeks may not be of much added clinical value, given the close monitoring of FGR pregnancies, particularly in the weeks leading up to birth. These clinical patient management schemes offer a much tighter range of potential and optimal delivery dates. Nonetheless, the purpose of our regressors is not to supplant current delivery prognosis practices, but to aid in providing tailored patient assessments of severity, maximising information extracted from MRI scans, not currently considered routine clinical practice (i.e. model fitting techniques and organ comparison assessments). 

From this, we demonstrate the ability of our method to provide insights into how fetal organs are affected in FGR, using this information to establish optimal delivery time within a two week range, which in future work may be of use to establish which pregnancies must be closely monitored. While we expect more severe cases to require early delivery, we do make important assumptions for these predictions, namely that all cases were delivered using the exact same criteria (when in practice patient view may also have influenced delivery choices), and that the appropriate and optimal clinical decisions were made, which is not unreasonable considering all our cases are from a specialised FGR unit. 

For this reason, we also investigated baby weight as a postnatal severity metric. We obtained optimal results for this metric, which demonstrates that fetal organ features such as perfusion are closely related to appropriate fetal growth, as determined by postnatal weight. 

The ResNet prediction results in Table \ref{tbl:cnn_results} concluded a much higher RMSE compared to our simpler logistic regression model. There are many potential reasons for this, such as the amount of noise in our CNN input data. Another evident reason for our poorer deep learning results is our small sample size, which, although we employed augmentation techniques, may still be insufficient to reliable train a CNN. Nonetheless, we obtained much closer results to our linear regressors for baby weight ResNet predictions. The fact that we included MRI scan data as our first channel may play a role in this, as baby weight is closely related to fetal size, which may be assessed from this first channel. Another factor to consider is the 6mm slice thickness of the scans being of a comparable size to the fetal organs. The structures of interest, such as the signal intensities of small vascular features and smaller tissue compartments (for instance in the fetal kidney), may have been susceptible to partial volume averaging compared to the brain, which is a bigger structure in comparison. However, our multi-compartment modelling takes this effect into account to some degree by attributing the signal from a single large voxel to different tissues.

Our deep learning method demonstrates how our organ model fitting maps contain spatial and intensity information which may be efficiently retrievable via CNNs, and presents potential to aid in providing condition information. Future work could test this directly with the current dataset by skipping the model fitting step. But there would be a resulting trade-off between interpretation (from validated MRI physiological models) and clinical predictivity (where ML techniques are a relative black-box for accurate prediction in absence of interpretability).

Our method proposed in this preliminary evaluation must be refined before translation to a clinical environment, but it may serve as a guide on condition severity. In practice though, this tool would also be used in conjunction with a wide range of information and existing biomarkers, including ultrasound data on fetal size, and maternal and fetal Doppler analysis of vascular resistance, which we have not included so far in this work. The ML analysis on these results supports the potential use of these parametric biomarkers in measuring FGR and providing an estimate of severity, including an indication of the likely GA at delivery. In addition to these biomarkers, future work could systematically include volumetric data on the brain, lungs, liver, and placenta to better enhance the ML models. However, the data was unregistered, did not use 3D reconstruction and would require direct comparisons to pre-published normative curves to know how lung/organ volume changes with gestation to incorporate fully. It is also important to note that the method employed assumes the delivery time of each subject was optimal, which although extracted from an early-onset clinic with specialised treatment, this may not be always the case, inducing biases. 
\par

The deep learning extension implemented to target this regression problem showcases potential avenues for future work with this type of voxelwise organ model fitted maps. These maps contain important spatial information, which proved useful to assess postnatal baby weight. Future work on deep learning should focus on appropriately selecting the input channel features, by conducting detailed assessments on the level of noise against information quality and significance. 

\par

Analysis on parameter correlations indicated that as the perfusion fraction in the liver and placenta decreased, the more severely growth-restricted the FGR fetuses were. This corroborated our initial hypotheses for selecting the fetal liver and placenta as severely-affected organs in FGR, with SNR perhaps too low and variability too high to observe differences in the fetal brain and lung. However, further work is needed to refine the analysis of the signals from these organs to better study the impact of FGR.
\par

Moreover, reliance of ML models on `Big Data' (\cite{wang2016machine}), motivates the need for a larger dataset, or data augmentation techniques to improve model performance and reduce generalisation error. Increased data availability could enable deep learning models, such as CNNs, which show potential for large-scale diagnosis improvement (\cite{yadav2019deep}), compared to traditional ML models. Our dataset of 24 subjects limits the results and conclusions from being generalised to the population. But this was not the purpose of the study. Rather, we sought to investigate the concepts and statistical methods employed in this paper. Future work could extend the methods to additional pregnancy complications to diagnose not only, FGR and non-FGR, but also the presence of other pregnancy conditions.
\par

\section{Conclusion}
In this proof-of-concept we proposed an approach to automate diagnosis of FGR using parameters extracted from the fetal liver and placenta, supported by the application of texture analysis. This preliminary investigation has demonstrated the potential of the models in assessing vascular properties of highly-perfused fetal organs, determined by multi-compartmental model fitting techniques. The placenta and fetal liver were prominent organs in identifying FGR fetuses, with key parametric features indicating a reduced perfusion, oxygenation and fetal capillary blood motion in these organs.

Our results prove that applying IVIM-based models on organs segmented from MRI scans generates features which are descriptive of FGR, i.e. potential biomarkers, enabling to construct simple machine learning models to predict diagnosis and offer insights into severity of the condition. The detailed voxel-level nature of our maps additionally enables deep learning experiments for condition severity assessments.

We validated our methodology on twenty-three FGR and control cases, achieving particularly optimal results for diagnosis classification. Our research exemplifies how ML models can be incorporated into the diagnostic workflow, as well as its potential to indicate severity of the condition. 
Future work into multi-organ fetal analysis will extend these techniques to other placental complications into a larger-scale study, using more complex ML and deep learning models.


\acks{This research was supported by the Wellcome Trust (210182/Z/18/Z, 101957/Z/13/Z, 203148/Z/16/Z and Wellcome Trust/EPSRC NS/A000027/1) and the Radiological Research Trust. The funders had no direction in the study design, data collection, data analysis, manuscript preparation or publication decision. We would like to thank Dr Magda Sokolska and Dr David Atkinson for their invaluable support and advice for the data collection in this work.}

%
\ethics{The work follows appropriate ethical standards in conducting research and writing the manuscript, following all applicable laws and regulations regarding treatment of animals or human subjects.}

\coi{We have no conflicts of interest to report.}

\bibliography{melba2022fetalorgans}

\newpage
\appendix
\section{Voxelwise Feature Importances}
\label{sec: voxelwise_appendix}

A total of 345 voxelwise measurements were extracted from the model fitting. Table \ref{tbl:voxel_significance_appendix} displays 50 of the features in order of feature importance in predicting an FGR diagnosis. This is an extended version of Table \ref{tbl:voxel_significance}.

\begin{tiny}

\begin{longtable}[c]{|p{1.5cm}|p{1.5cm}|p{1cm}|p{1.5cm}|p{1cm}|p{1.5cm}|p{1.5cm}|}
    \hline 
        \textbf{Model Fitting Technique} & \textbf{Parameter} & \textbf{Average Metric} & \textbf{Pairwise Group Comparison} & \textbf{Organ} & \textbf{T Statistic} & \textbf{P-Value} \\ \hline\hline
        Dependent IVIM & D* & Mean & Control vs FGR & Placenta & -4.597300242 & 0.00015589 \\ \hline
        Extended 2xT2 Depedent IVIM & D* & Mean & Control vs FGR & Placenta & -4.560436097 & 0.000170214 \\ \hline
        DECIDE Model (Voxelwise Measurements) & D* & Mean & Control vs FGR & Placenta & -4.205788361 & 0.00039723 \\ \hline
        Extended 2xT2 Depedent IVIM & Perfusion Fraction & Min & Control vs FGR & Placenta & 3.725183003 & 0.001250966 \\ \hline
        Extended 2xT2 Depedent IVIM & Perfusion Fraction & Mode & Control vs FGR & Placenta & 3.725183003 & 0.001250966 \\ \hline
        Standard IVIM & Perfusion Fraction & Median & Control vs FGR & Liver & 3.624757118 & 0.001587669 \\ \hline
        Dependent IVIM & T2 & Min & Control vs FGR & Placenta & 3.463092031 & 0.002326109 \\ \hline
        Extended 2xT2 Depedent IVIM & Perfusion Fraction & Median & Control vs FGR & Placenta & 3.27041186 & 0.003653498 \\ \hline
        Dependent IVIM & Perfusion Fraction & Min & Control vs FGR & Placenta & 3.249455242 & 0.003836258 \\ \hline
        Dependent IVIM & Perfusion Fraction & Mode & Control vs FGR & Placenta & 3.249455242 & 0.003836258 \\ \hline
        Standard IVIM & D* & Mean & Control vs FGR & Placenta & -3.155410162 & 0.004771861 \\ \hline
        T2 Fitting & T2 & Mode & Control vs FGR & Placenta & 3.076054116 & 0.005730308 \\ \hline
        T2 Fitting & T2 & Min & Control vs FGR & Placenta & 3.076054116 & 0.005730308 \\ \hline
        Dependent IVIM & Perfusion Fraction & Max & Control vs FGR & Placenta & -2.908584282 & 0.008399742 \\ \hline
        DECIDE Model (Voxelwise Measurements) & Perfusion Fraction & Mean & Control vs FGR & Placenta & -2.860182788 & 0.009371321 \\ \hline
        Extended 2xT2 Depedent IVIM & Perfusion Fraction & Mode & Control vs FGR & Brain & -2.846049894 & 0.010722475 \\ \hline
        Extended 2xT2 Depedent IVIM & Perfusion Fraction & Max & Control vs FGR & Brain & -2.846049894 & 0.010722475 \\ \hline
        Dependent IVIM & D* & Min & Control vs FGR & Placenta & 2.749012922 & 0.012025317 \\ \hline
        Dependent IVIM & Perfusion Fraction & Median & Control vs FGR & Placenta & 2.746991901 & 0.012079621 \\ \hline
        Extended 2xT2 Depedent IVIM & Fetal Blood T2 & Min & Control vs FGR & Placenta & 2.661186891 & 0.014612077 \\ \hline
        ADC Fitting & ADC & Mode & Control vs FGR & Placenta & 2.60516097 & 0.016528088 \\ \hline
        ADC Fitting & ADC & Min & Control vs FGR & Placenta & 2.60516097 & 0.016528088 \\ \hline
        DECIDE Model (Voxelwise Measurements) & Perfusion Fraction & Mode & Control vs FGR & Placenta & 2.602142163 & 0.01663777 \\ \hline
        DECIDE Model (Voxelwise Measurements) & Perfusion Fraction & Min & Control vs FGR & Placenta & 2.602142163 & 0.01663777 \\ \hline
        Dependent IVIM & Perfusion Fraction & Mean & Control vs FGR & Placenta & 2.589134143 & 0.017118271 \\ \hline
        Standard IVIM & Perfusion Fraction & Mean & Control vs FGR & Liver & 2.539211902 & 0.019086068 \\ \hline
        Standard IVIM & S0 & Median & Control vs FGR & Placenta & 2.502100007 & 0.020684193 \\ \hline
        T2 Fitting & T2 & Median & Control vs FGR & Placenta & 2.484716827 & 0.021475009 \\ \hline
        Standard IVIM & S0 & Mode & Control vs FGR & Placenta & 2.467133972 & 0.022303541 \\ \hline
        Standard IVIM & S0 & Min & Control vs FGR & Placenta & 2.467133972 & 0.022303541 \\ \hline
        Standard IVIM & Tissue T2 & Mean & Control vs FGR & Placenta & 2.456622652 & 0.022812982 \\ \hline
        Extended 2xT2 Depedent IVIM & Perfusion Fraction & Median & Control vs FGR & Placenta & 2.415973093 & 0.024886753 \\ \hline
        Standard IVIM & Perfusion Fraction & Max & Control vs FGR & Placenta & -2.412729204 & 0.02505957 \\ \hline
        Extended 2xT2 Depedent IVIM & D* & Mode & Control vs FGR & Liver & -2.402194517 & 0.025628509 \\ \hline
        Dependent IVIM & D* & Max & Control vs FGR & Placenta & -2.271942855 & 0.033722187 \\ \hline
        Dependent IVIM & T2 & Max & Control vs FGR & Liver & 2.256543771 & 0.034820546 \\ \hline
        Dependent IVIM & T2 & Median & Control vs FGR & Placenta & 2.248766366 & 0.035387639 \\ \hline
        DECIDE Model (Voxelwise Measurements) & D* & Mode & Control vs FGR & Placenta & 2.223360922 & 0.037299443 \\ \hline
        Dependent IVIM & D* & Min & Control vs FGR & Placenta & 2.223360922 & 0.037299443 \\ \hline
        ADC Fitting & ADC & Mode & Control vs FGR & Lung & -2.201659535 & 0.039006645 \\ \hline
        ADC Fitting & ADC & Min & Control vs FGR & Lung & -2.201659535 & 0.039006645 \\ \hline
        Extended 2xT2 Depedent IVIM & Tissue T2 & Min & Control vs FGR & Brain & -2.179361415 & 0.042831566 \\ \hline
        T2 Fitting & S0 & Mode & Control vs FGR & Placenta & 2.138821399 & 0.044358479 \\ \hline
        T2 Fitting & S0 & Min & Control vs FGR & Placenta & 2.138821399 & 0.044358479 \\ \hline
        ADC Fitting & S0 & Median & Control vs FGR & Placenta & 2.127953943 & 0.04534878 \\ \hline
        DECIDE Model (Voxelwise Measurements) & Maternal Blood Volume & Median & Control vs FGR & Placenta & 2.08157448 & 0.049803299 \\ \hline
        Standard IVIM & Perfusion Fraction & Median & Control vs FGR & Placenta & 2.063763284 & 0.051616017 \\ \hline
        Extended 2xT2 Depedent IVIM & S0 & Mean & Control vs FGR & Brain & -2.057178836 & 0.054456083 \\ \hline
        Dependent IVIM & S0 & Mode & Control vs FGR & Placenta & 2.003124411 & 0.058239086 \\ \hline
        Dependent IVIM & S0 & Min & Control vs FGR & Placenta & 2.003124411 & 0.058239086 \\ \hline
    \caption{Hierarchy of parameter feature importances of the voxelwise measurements (top 50 features).}\label{tbl:voxel_significance_appendix}
\end{longtable}
\end{tiny}

\clearpage
\section{Haralick Feature Importances}
\label{sec: haralick_appendix}

A total of 172 Haralick features were extracted, 50 of which are displayed in Table \ref{tbl:haralick_significance_appendix} in order of feature importance.

\begin{tiny}
\begin{longtable}[c]{|p{1.5cm}|p{1.2cm}|p{1.5cm}|p{1.5cm}|p{1cm}|p{1.5cm}|p{1.5cm}|}
    \hline
        \textbf{Model Fitting Technique} & \textbf{Parameter} & \textbf{Haralick Feature} & \textbf{Pairwise Group Comparison} & \textbf{Organ} & \textbf{T Statistic} & \textbf{P-Value} \\ \hline\hline
        Extended 2xT2 Dependent IVIM & D* & Mean Variance & Control vs FGR & Placenta & 3.85713275 & 0.000913732 \\ \hline
        Extended 2xT2 Dependent IVIM & D* & Mean Contrast & Control vs FGR & Placenta & -3.59999694 & 0.001683568 \\ \hline
        Extended 2xT2 Dependent IVIM & D* & Mean Energy & Control vs FGR & Placenta & -3.52882897 & 0.001992104 \\ \hline
        Extended 2xT2 Dependent IVIM & D* & Mean Energy & Control vs FGR & Placenta & 3.47784754 & 0.002246655 \\ \hline
        Extended 2xT2 Dependent IVIM & D* & Max Correlation & Control vs FGR & Placenta & 3.314572951 & 0.003295715 \\ \hline
        T2 Fitting & Perfusion Fraction & Mean Correlation & Control vs FGR & Placenta & -3.24466242 & 0.003879289 \\ \hline
        Extended 2xT2 Dependent IVIM & D* & Max Homogeneity & Control vs FGR & Placenta & -3.01289081 & 0.006623681 \\ \hline
        T2 Fitting & Perfusion Fraction & Mean Contrast & Control vs FGR & Placenta & 2.8780617 & 0.00900053 \\ \hline
        Standard IVIM & Perfusion Fraction & Max Contrast & Control vs FGR & Liver & 2.843764223 & 0.00972465 \\ \hline
        Extended 2xT2 Dependent IVIM & D* & Mean Homogeneity & Control vs FGR & Placenta & -2.7464043 & 0.012095454 \\ \hline
        b=0 Volume & - & Max correlation & Control vs FGR & Liver & -2.69457234 & 0.01357218 \\ \hline
        Standard IVIM & Perfusion Fraction & Mean Entropy & Control vs FGR & Liver & 2.672682492 & 0.014245788 \\ \hline
        Extended 2xT2 Dependent IVIM & Perfusion Fraction & Max Energy & Control vs FGR & Liver & -2.65258805 & 0.014891863 \\ \hline
        Extended 2xT2 Dependent IVIM & Perfusion Fraction & Mean Entropy & Control vs FGR & Liver & 2.63659631 & 0.015425697 \\ \hline
        Standard IVIM & Perfusion Fraction & Max Entropy & Control vs FGR & Liver & 2.629661922 & 0.015662746 \\ \hline
        b=0 Volume & - & Max Energy & Control vs FGR & Lung & -2.62851088 & 0.015702425 \\ \hline
        ADC Fitting & Perfusion Fraction & Mean Variance & Control vs FGR & Lung & -2.2586354 & 0.015794044 \\ \hline
        Standard IVIM & Perfusion Fraction & Mean Variance & Control vs FGR & Liver & -2.61635759 & 0.016127196 \\ \hline
        Standard IVIM & Perfusion Fraction & Mean Contrast & Control vs FGR & Liver & 2.589841409 & 0.017091814 \\ \hline
        Extended 2xT2 Dependent IVIM & Perfusion Fraction & Mean Energy & Control vs FGR & Liver & -2.56185367 & 0.018168659 \\ \hline
        Extended 2xT2 Dependent IVIM & Perfusion Fraction & Max Entropy & Control vs FGR & Liver & 2.547927974 & 0.018727882 \\ \hline
        b=0 Volume & - & Mean Entropy & Control vs FGR & Lung & -2.54682773 & 0.018772746 \\ \hline
        b=0 Volume & - & Mean Variance & Control vs FGR & Lung & -2.5463024 & 0.018794203 \\ \hline
        Standard IVIM & Perfusion Fraction & Max Energy & Control vs FGR & Liver & -2.54431325 & 0.018875657 \\ \hline
        ADC Fitting & Perfusion Fraction & Mean Energy & Control vs FGR & Lung & -2.54246834 & 0.0189515 \\ \hline
        Standard IVIM & Perfusion Fraction & Mean Energy & Control vs FGR & Liver & -2.52318667 & 0.019761426 \\ \hline
        ADC Fitting & Perfusion Fraction & Mean Entropy & Control vs FGR & Lung & 2.519857262 & 0.019904517 \\ \hline
        b=0 Volume & - & Mean Entropy & Control vs FGR & Lung & 2.514269872 & 0.020146831 \\ \hline
        T2 Fitting & Perfusion Fraction & Max Homogeneity & Control vs FGR & Placenta & 2.497234523 & 0.020902754 \\ \hline
        ADC Fitting & Perfusion Fraction & Mean Variance & Control vs FGR & Lung & -2.4836657 & 0.021523722 \\ \hline
        Extended 2xT2 Dependent IVIM & Perfusion Fraction & Mean Variance & Control vs FGR & Liver & -2.44831811 & 0.023223109 \\ \hline
        b=0 Volume & - & Mean Correlation & Control vs FGR & Placenta & -2.4461168 & 0.023332967 \\ \hline
        Standard IVIM & Perfusion Fraction & Max Correlation & Control vs FGR & Liver & -2.42840779 & 0.024234499 \\ \hline
        b=0 Volume & - & Max Homogeneity & Control vs FGR & Placenta & 2.413012066 & 0.025044456 \\ \hline
        b=0 Volume & - & Max Energy & Control vs FGR & Liver & -2.37816262 & 0.02697148 \\ \hline
        T2 Fitting & Perfusion Fraction & Max Correlation & Control vs FGR & Placenta & -2.37038089 & 0.027420117 \\ \hline
        b=0 Volume & - & Mean Homogeneity & Control vs FGR & Liver & 2.334287019 & 0.02959262 \\ \hline
        b=0 Volume & - & Max Contrast & Control vs FGR & Liver & 2.281290927 & 0.033071008 \\ \hline
        Extended 2xT2 Dependent IVIM & Perfusion Fraction & Max Contrast & Control vs FGR & Liver & 2.279938553 & 0.033164494 \\ \hline
        b=0 Volume & - & Max Variance & Control vs FGR & Lung & -2.27647378 & 0.033405112 \\ \hline
        b=0 Volume & - & Mean Entropy & Control vs FGR & Liver & 2.267311575 & 0.034049134 \\ \hline
        b=0 Volume & - & Mean Energy & Control vs FGR & Liver & -2.25902498 & 0.034641385 \\ \hline
        b=0 Volume & - & Mean Contrast & Control vs FGR & Liver & 2.25382319 & 0.035017302 \\ \hline
        Extended 2xT2 Dependent IVIM & Perfusion Fraction & Mean Contrast & Control vs FGR & Liver & 2.243927048 & 0.035744752 \\ \hline
        T2 Fitting & Perfusion Fraction & Mean Homogeneity & Control vs FGR & Placenta & 2.243681655 & 0.035762948 \\ \hline
        b=0 Volume & - & Max Entropy & Control vs FGR & Liver & 2.238043495 & 0.036183359 \\ \hline
        ADC Fitting & Perfusion Fraction & Max Variance & Control vs FGR & Lung & -2.22990394 & 0.036798255 \\ \hline
        b=0 Volume & - & Mean Variance & Control vs FGR & Liver & -2.20345811 & 0.038862496 \\ \hline
        b=0 Volume & - & Max Entropy & Control vs FGR & Lung & 2.200623685 & 0.039096886 \\ \hline
        ADC Fitting & Perfusion Fraction & Max Entropy & Control vs FGR & Lung & 2.195186635 & 0.039529463 \\ \hline
    \caption{Hierarchy of most significant Haralick features across parameter maps and organs (top 50 features).}\label{tbl:haralick_significance_appendix}
\end{longtable}
\end{tiny}

\end{document}